# Wideband photonic interference cancellation based on free space optical communication

YANG QI, AND BEN WU[*]

*Department of Electrical and Computer Engineering, Rowan University, Glassboro, New Jersey 08028 USA*
*[*]wub@rowan.edu*

**Abstract:** We propose and experimentally demonstrate an interference management system that removes wideband wireless interference by using photonic signal processing and free space optical communication. The receiver separates radio frequency interferences by upconverting the mixed signals to optical frequencies and processing the signals with the photonic circuits. Signals with GHz bandwidth are processed and separated in real-time. The reference signals for interference cancellation are transmitted in a free space optical communication link, which provides large bandwidth for multi-band operation and accelerates the mixed signal separation process by reducing the dimensions of the un-known mixing matrix. Experimental results show that the system achieves 30dB real-time cancellation depth with over 6GHz bandwidth. Multiple radio frequency bands can be processed at the same time with a single system. In addition, multiple radio frequency bands can be processed at the same time with a single system.

## 1. Introduction

The emerging as well as the evolving communication technologies rely on using increasingly higher radio frequencies (RF) for larger communication bandwidth. The unprecedented channel capacity enabled by using higher carrier frequencies provides high-speed communication for commercial and active users. However, the spectrum for passive users and some essential active users depends on the intrinsic properties of the scientific object to be observed and cannot be switched to higher frequencies. For example, as a passive user, a radio telescope defines its spectrum window based on the radiation of celestial objects and the spectrum window cannot be changed based on the availability of spectrum [1–3]. As another example, weather radar, as an essential active user, selects the spectrum window based on the spectral response of the meteorological phenomena to be observed [4,5]. More importantly, higher frequency signal carriers require line-of-sight (LOS) transmission, which limits their deployment and requires hybrid systems that use both existing low frequencies as well as undeveloped high frequencies [6–8]. Therefore, the dramatic growth of commercial and active users generates increasingly more interference in both high and low frequencies for passive users. To enable the coexistence of passive and active systems, interference management system that operates in a wideband of RF spectrum is needed.

Photonic signal processing has been demonstrated to process wideband RF signals in real-time and has been applied to spectral filtering [9–13], blind source separation (BSS) [14–18], and removement self-interference for full duplexing [19–22]. The photonic system processes the wideband signals by modulating RF spectrum on optical carriers. Since the optical frequencies are 4 to 5 orders of magnitude larger than the RF frequencies, a wideband in RF frequencies can be treated as a narrow band in optical frequencies. With bandwidth in THz range, a photonic system can separate signals for different RF bands, from sub-6GHz to millimeter waves. Such an advantage is desired with the growing needs of software define radio, where a single hardware is able to support a wide range of RF protocols [23].

In this paper, we make use of the photonic signal processing method to remove the interference generated from a multi-user system. The cancellation system functions as spectral trading method between high and low frequencies. Fig. 1 (b) shows the "trade" by a logarithmic scale spectrum. A "narrow" band at higher frequency can trade n copies of "wide" band at lower frequency. With such a trade, the congestion problem at lower frequencies (RF) is solved without consuming much of the spectrum resources at higher frequencies.

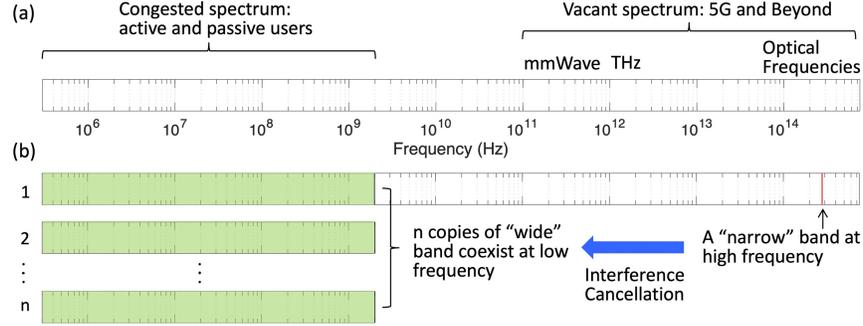

Fig. 1 Comparison of (a) a single RF spectrum (b) Coexistence of multiple copies RF spectra

The trading method is enabled by sending a reference signal from the interference generator, and modulating the reference signal on high frequency carriers. With the traded copies of spectrum, multiple transmitter / receiver pairs can coexist. Fig. 2 shows the coexistence of two transmitter / receiver pairs. The receiver Rx1 receives SOI $s_{soi}$ from transmitter Tx1 and interference $s_{int}$ from Tx2. SOI and interference signals are in the same RF band. Tx2 is closer to Rx1 than Tx1. Rx1 is within LOS transmission range of Tx2, and is not within LOS transmission of Tx1. With interference, the communication between Tx1 and Rx1 cannot simply switch to higher frequencies (millimeter waves and free space optical communication), because the LOS transmission is blocked by obstacles between Tx1 and Rx1. This is a common case that requires interference cancellation, where the receiver is closer to the interference source than the SOI source. To remove the interference from Tx2, a reference signal is sent from Tx2 to Rx1 in a free space optical communication link. Rx1 cancels the interference with the reference signal. The free space optical communication link can be seamlessly integrated with the photonic interference cancellation system, and both the free space optical communication link and the photonic interference cancellation system operate in a wide range of RF frequencies.

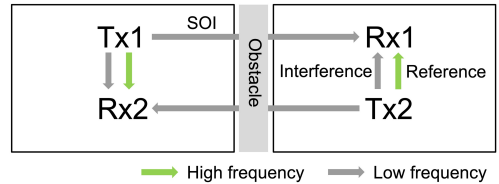

Fig. 2 Interference cancellation enabled by high frequency carriers (optical frequencies). The obstacle blocks the LOS transmission of high frequency signal, and low frequency signal can pass through the obstacle. (SOI: signal of interest)

## 2. Principle and experimental setup

This section discusses the principle of interference cancellation and the experimental setup of the photonic cancellation system. The experimental setup in section 2.1 shows that FSO channel is compatible with the photonic signal processing system for interference cancellation. Section 2.2 shows that the separation process can be simplified with the FSO link.

*2.1 Experimental setup*

The experimental setup demonstrated the functions of Tx1 SOI transmitter, Tx2 interference generator, and Rx1 SOI receiver with interference cancellation (Fig. 2). SOI is generated from a vector signal generator (Keysight M9381A PXIe). The modulation formats of SOI include QPSK, 16QAM, 64QAM, and 256QAM, and carrier frequencies of SOI range from 500MHz to 6GHz. Interference signal is generated from signal generators HP ESG-D4000A and Gigatronics 7100. The interference signal is Gaussian noise frequency modulated on the carriers. The bandwidth of interference overlaps with the bandwidth of SOI. The FSO channel modulates the RF reference signal on a distributed feedback laser with wavelength of 1544nm and has a transmission distance of 1m. The receiver Rx1 modulates the mixed RF signals on another laser carrier with wavelength of 1560nm. By controlling the bias of the intensity modulator (MOD2), the interference signal from the RF channel can be either added with the reference signal from the FSO channel or subtracted from the reference signal. The subtraction achieves the interference cancellation function. The tunable delays and tunable attenuators at the receiver match the phases and amplitudes between the interference and the reference signals.

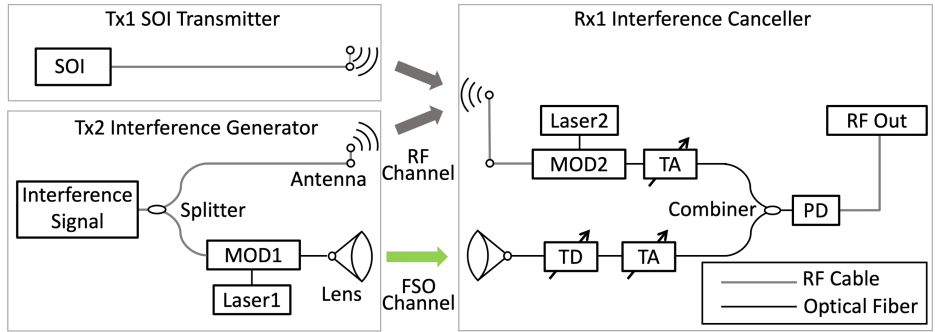

Fig. 3 Experimental setup of the interference cancellation system. (MOD1 and MOD2: Optical intensity modulators, FSO: free space optical communication, TD: optical tunable delay, TA: optical tunable attenuator, PD: photodiode)

The matching condition is achieved by photonic BSS method. By controlling the weights of the mixed signals with tunable attenuators and measuring the statistical information of the mixed signals at different weights, the de-mixing matrix is calculated by the BSS algorithm. The FSO channel is seamlessly integrated with the photonic BSS system without converting to electrical signals. Using the FSO channel to provide reference signals can be treated as a special case for BSS. The reference signals only include interference and do not include SOI, which simplifies the mixing matrix.

*2.2 Photonic blind source separation*

Photonic BSS has been demonstrated to separate wideband interference from SOI without pre-known information about the mixed signals [14]. Photonic BSS processes the mixed signal by modulating the RF signals on optical carriers. Since wideband RF signals (GHz bandwidth) can be treated as narrow band signals at optical frequencies, photonic BSS processes wideband RF signals in real-time and deep cancellation ratio. The FSO link is compatible with the photonic BSS system. The reference signal in the FSO channel is transmitted on optical carriers and can be directly processed with the tunable weight and delay of the photonic BSS system. More importantly, the reference signal simplifies the BSS algorithm and accelerates the system response for finding the de-mixing matrix.

The photonic BSS system processes signals from a MIMO receiver. The mixed signals are represented by:

$$\boldsymbol{X} = \boldsymbol{AS} \text{ or, } \begin{bmatrix} x_1 \\ x_2 \end{bmatrix} = \begin{bmatrix} a_{11} & a_{12} \\ a_{21} & a_{22} \end{bmatrix} \begin{bmatrix} s_{soi} \\ s_{int} \end{bmatrix} \quad (1)$$

$A$ is the mixing matrix, and $x_1$ and $x_2$ are mixed signals received from MIMO antennas. $s_{soi}$ and $s_{int}$ are SOI and interference respectively. By using the FSO channel to transmit the reference signal, one of the MIMO antennas is replaced by the FSO channel, and therefore Eq. 1 changes to:

$$\begin{bmatrix} r_L \\ r_H \end{bmatrix} = \begin{bmatrix} a_{11} & a_{12} \\ 0 & a_{22} \end{bmatrix} \begin{bmatrix} s_{soi} \\ s_{int} \end{bmatrix} \quad (2)$$

Where $r_L$ is the signal on low frequency carriers (congested RF frequencies received from antenna) and is a mixture of $s_{soi}$ and $s_{int}$. $r_H$ is the reference signal from high frequency carries (optical frequencies). Since the interference generator (Fig. 2) only send a copy of the interference signal, $r_H$ only includes interference $s_{int}$, and $a_{21} = 0$. Without the reference signals, to find the de-mixing matrix $A^{-1}$ is challenging and time-consuming for the BSS. With the pre-known parameter of $a_{21} = 0$, both the principal component analysis and the independent component analysis of the BSS process are simplified.

## 3. Experimental results and analysis

### 3.1 Demonstration of interference cancellation

The cancellation system is firstly demonstrated with interference and SOI both at 2.4GHz, a most widely used band for both passive and active users. Fig. 4 (a) shows the constellation diagram of the SOI without interference introduced. The SOI used in this experiment is QPSK signal with symbol rate of 5MBd. The QPSK is chosen to be SOI, so the patterns of SOI can be easily identified. Without loss of generality, and the separation system also applies to other modulation formats. The signal is received by Rx1 in Fig. 3. Fig. 4 (b) shows the spectrum of the received signal, which has a carrier frequency of 2.4GHz and bandwidth of 5MHz. The interference in this experiment is random Gaussian noise that is frequency modulated on a 2.4GHz carrier frequency. The peak-to-peak frequency deviation is 80MHz. The interference is 18dB stronger than the SOI. Fig. 4 (c) shows constellation diagram of the received signal with both interference and SOI and without the cancellation system working. The constellation diagram is completely noisy and the pattern of the QPSK cannot be identified. Fig. 4 (d) shows the spectrum of the mixed signal. Since the interference is 5dB stronger and also has a wider bandwidth than the SOI, the spectrum of the interference completely covers the spectrum of the SOI and the spectrum pattern of the SOI cannot be observed.

Fig. 4 (e) shows that by using the photonic interference cancellation system and the FSO communication link, the wideband interference is effectively cancelled, and a clear constellation diagram for SOI is recovered. The spectrum of the recovered signal also shows a cancellation of the interference spectrum. Both the constellation diagram and the spectrum of the recovered SOI (Figs. 4 (e) and (f)) match with the constellation diagram and the spectrum without interference introduced (Figs. 4 (a) and (b)).

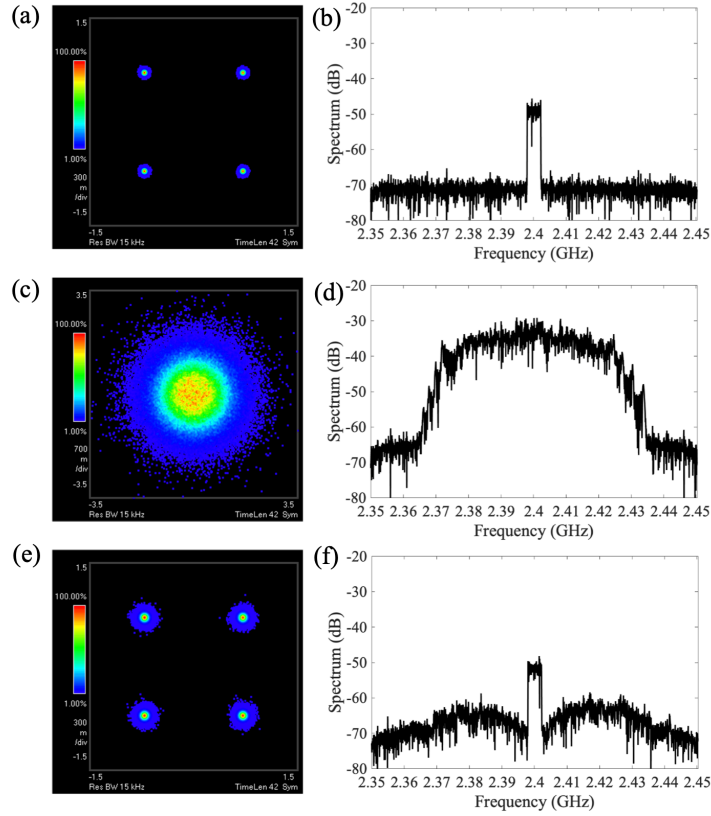

Fig. 4 Received signals with and without interference cancellation (a) Constellation diagram of the received signal without interference introduced (b) Spectrum of the received signal without interference introduced (c) Constellation diagram of the received signal with interference introduced (d) Spectrum of the received signal with interference introduced (e) Constellation diagram of the received signal with interference and interference cancellation (f) Spectrum of the received signal with interference and interference cancellation.

The comparison of the spectrum with and without interference cancellation shows that the photonic system suppresses the interference by over 30dB (Fig. 5 (a)). To quantitatively evaluate the received SOI, the error vector magnitude (EVM) of the received constellation diagram is measured at different interference to SOI ratios (Fig. 5 (b)). The interference to SOI ratio is defined as the ratio of spectral density between the interference and SOI at 2.4GHz. Without using the photonic interference cancellation system and the FSO link, the EVM of the received signals increase from 5% to over 70% when the ratio of the interference to SOI increases from -25dB to 18dB. With over 70% EVM, the received constellation diagram is noisy without patterns of the SOI. Fig. 4 (c) and the black line in Fig. 5 (a) correspond to the black arrow in Fig. 5 (b). By using the photonic interference cancellation system and the FSO link, the EVM of the received signals is less than 15% when the ratio of the interference to SOI increases from -25dB to 18dB. The received constellation diagram shows clear patterns of QPSK signals, and Fig. 4 (e) and the red line in Fig. 5 (a) correspond to the red arrow in Fig. 5 (b). The measurement of EVM shows that the system effectively recovers the SOI when the ratio of interference to SOI changes in a wide range.

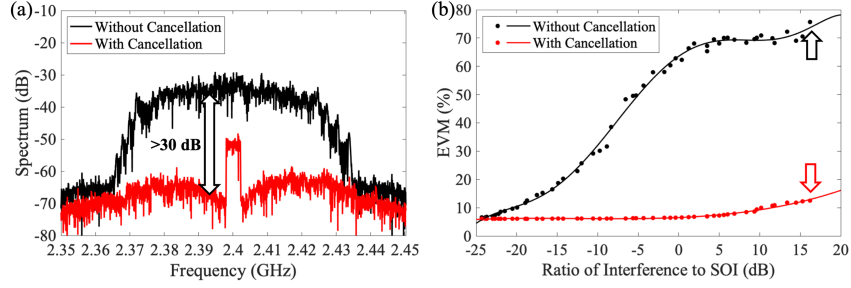

Fig. 5 (a) Comparison of the received spectrum with and without interference cancellation (b) Comparison of error vector magnitude (EVM) with and without interference cancellation.

*3.2 Signal separation for different modulation formats*

In addition to QPSK signal, the cancellation system is tested with SOI with other modulation formats, including 16QAM, 64QAM and 256QAM. Figs. 6 (a), (c), and (e) show the clear constellation diagrams with the cancellation system, and Figs. 6 (b), (d), and (f) show the noisy constellation diagrams without the cancellation system. All the measurements in Fig. 6 use 2.4GHz carrier frequency and the interference 9dB stronger than the SOI. Since the photonic system separates the interference from the SOI in an analog way, the interference cancellation is performed without pre-known information of modulation format of either the interference or the SOI. The system is compatible with any types of modulation format. The over 30dB cancellation ratio enables the system to recover SOI with complex modulation format, such as the 256QAM in Fig. 6 (e).

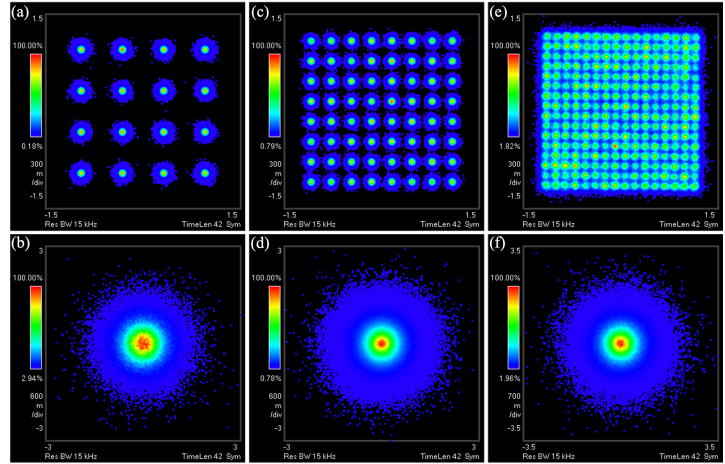

Fig. 6 Constellation diagrams when the SOI is 16QAM: (a) and (b); 64QAM: (c) and (d); and 256QAM: (e) and (f). (a), (c), and (e) are recovered SOI with the cancellation system. (b), (d), and (f) are the received mixed signals without the cancellation system.

*3.3 Wideband spectral response of the cancellation system*

The interference cancellation system works in a wide spectral range without the needs of extra tuning. Fig. 7 shows the spectral response of the system measured with a network analyzer (Keysight E5063A). The black line shows the transmission spectrum of interference signal without cancellation, and the red line shows the transmission spectrum of interference signal with cancellation. The comparison between the black line and the red line shows that over 30dB of interference cancellation is achieved from direct current (DC) to 4GHz, and 20dB of interference cancellation is achieved from 4GHz to 6GHz. The drop of cancellation ratio at 4GHz to 6GHz is caused by the frequency response of the intensity modulator.

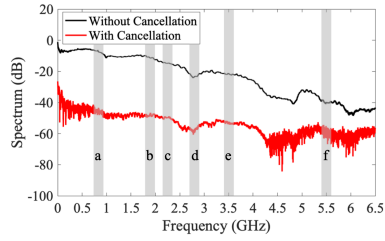

Fig. 7 Spectrum response of the system.

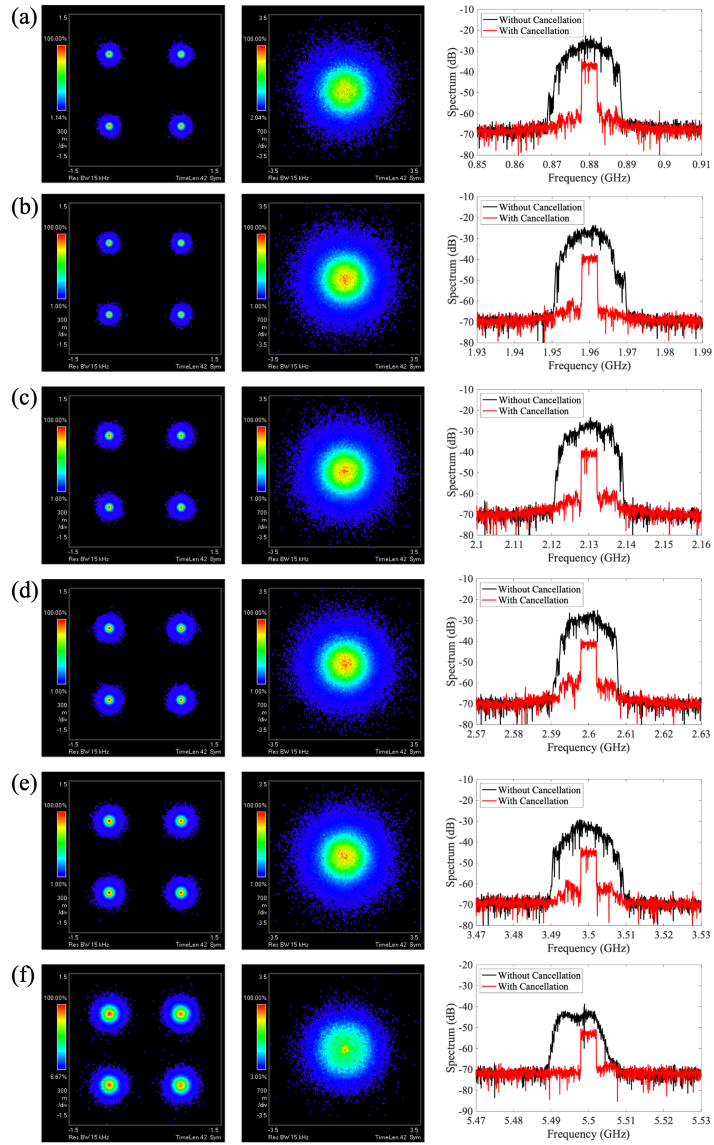

Fig. 8 Constellation diagrams and spectra of different communication protocols with a wide range of carrier frequencies.

Figs. 8 (a)-(f) show the interference cancellation for a wide range of communication protocols with frequencies range from 870 MHz to 5.51 GHz. The figures in the left column and the middle column are the constellation diagrams with and without cancellation, and the figures in the right column are the spectra of the received signals with and without cancellation. To eliminate the impact of antennas with limited bandwidth, the interference and SOI are sent from signal generators to receiver with RF cables, RF splitters and combiners. The spectral response of the system is determined by the optical intensity modulator that upconvert the RF signal to optical frequencies, and the photodetector that down convert optical signal to RF bands (Fig. 3). If the photonic system is treated as a black box for signal processing, the modulator and photodetector are the entrance and exit of the black box. The speeds of modulation and photodetection determine the signal bandwidths that flow into and out of the black box. Once the RF signal is upconverted to optical frequencies (193Thz) and within the black box, the bandwidth that the system can process (up to 5THz) is far beyond the maximum frequency of RF spectrum, or in another word, the photonic processer is nearly frequency independent in the RF range.

In this experiment, the bandwidth of intensity modulator is 9GHz, and bandwidth of the photodetector is 40GHz. The spectral response measured in Fig. 7 is determined by the spectral response of the intensity modulator. The most recent modulation and photodetection techniques achieve bandwidth of 100GHz [24], which not only covers the sub-6GHz RF frequencies for the current 4[th] generation communication, but also covers the millimeter wave communication for the 5[th] generation network.

### 3.4 Interference Cancellation in a Heterogeneous Network Environment

Experimental results in Sections 3.1 to 3.3 demonstrate interference cancellation for two transmitter / receiver pairs. In this section, the application of interference cancellation is scaled up to a heterogeneous wireless network. In such a network, the interference cancellation system (1) enables passive users' continuous access to the full RF spectrum (2) solves the unbalanced problem that large bandwidth is only available at places covered by 5G microcells, and spectrum congestion still exists at places not covered by microcells, and (3) most importantly, can be seamlessly integrated with the evolving 5G networks.

The RF spectrum is congested with multiple wireless systems with different protocols. To share the RF spectrum requires each user to be allocated to a different time slots, and each user works intermittently [25–27]. Fig. 9(a) shows a network with two active users A and B, e.g., cell phone users, and a passive user C, e.g., a radio telescope with in the same mobile cell. The green bars show that each user is allocated with 1/3 of the time slots. Compared with active users, the intermittent access to RF spectrum is more fatal to passive users. For active users, the SOI can be turned on and off based on the availability of the channels, and the intermittently available channel only affects the speed of communication and introduces latency. While for passive users, such as the radio telescope, the radiation of astronomical objects cannot be adjusted based on the channel availability. Therefore, important information may be missed when the channel is not available.

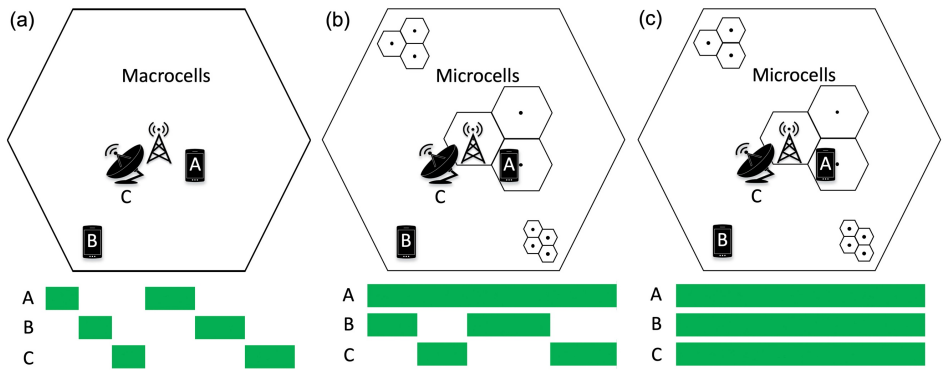

Fig. 9 Comparison of passive and active users' access to RF spectrum among (a) 4G macrocell (b) hybrid 4G macrocell and 5G microcell (c) hybrid 4G macrocell and 5G microcell with the interference cancellation system. The green bars show the time slots that the same RF spectrum is available for users A, B, and C.

This problem can be relieved but not completely solved by switching to higher frequencies. The 5G networks and beyond provide large bandwidth communication by using signal carrier with higher frequencies, such as 50-70GHz, millimeter waves, terahertz waves, and optical frequencies [28–33]. As shown in Fig. 9 (b), user A switches to higher frequencies, so it not only communicate with larger bandwidths, but also step out of the competition for spectrum at lower frequencies. However, the LOS transmission property of higher frequencies requires smaller cells to be built [34–36], and microcells may not be available in all the places. This is why a hybrid system is fundamental network infrastructure in 5G networks [37–40]. For a user that is not covered by the microcells, such as user B in Fig. 9 (b), it still competes for spectrum with the passive user C, and either generates interference to C or forces C to receive signal intermittently.

The interference cancellation system method solves the problem of receiving signals intermittently for passive users by providing reference signals carried by high frequency carriers to the passive users (radio telescope C). With interference cancellation enabled by the reference signals, the passive users can coexist with active user B at the same time slot. The interference cancellation system also leverages the unbalanced problem of the 5G network. There is a huge bandwidth difference between physical locations covered by 5G microcells and locations not covered. The limited coverage of microcells not only stems from the fact that 5G and undeveloped frequencies will be deployed step by step, but also because in many cases, receivers (such as user B in Fig. 9) are surrounded with obstacles, and cannot be reached by LOS transmission. As explained in Fig. 1, instead of simply having some active users to switch to higher frequencies, and left other active users in the congested RF spectrum, generating interference to the passive users, the cancellation system removes the interference, which equals to generate copies of RF spectrums.

Last but not the least, the hardware network infrastructure in Figs. 9(b) and 9(c) are very close to each other, which means the cancellation system can be seamlessly integrated with the 5G microcell networks with very low extra cost. In Section 3.1, the reference signals are sent with FSO channels, which is part of 5G network [41–43]. The reference can also be sent with other carrier frequencies of 5G network, including 50-70GHz, and millimeter waves, or carrier frequencies beyond 5G, such as terahertz waves. In particular, the passive users, such as the radio telescope stations, are immobile, which further simplifies the process of matching channel coefficients for interference cancellation.

## 4. Conclusion

We have proposed and experimentally demonstrated a photonic interference cancellation system that separates the interference from SOI based on the reference signal in a free space optical communication link. By using a free space optical communication link to carry the reference signal, the de-mixing matrix is simplified with reduced dimensions. By modulating the mixed signals on optical carriers, the system is able to process RF signals with 6GHz bandwidth. With the ultra-wide bandwidth, a variety of communication protocols, ranging from 870 MHz to 5.51 GHz can be processed by the same system without extra tuning. Experimental results show that 30dB cancellation is achieved in GHz range, which meet the needs of separating strong interference from weak SOI. As an analog signal processing method, the cancellation system separates the interference without pre-known information of the modulation formats of both the interference and SOI.

**Funding.** This research is supported by Rowan University startup grant and New Jersey Health Foundation (Grant # PC 77-21).

**Acknowledgments.** The authors would like to thank Taichu Shi, Department of Electrical Engineering, Rowan University, for help with lab and equipment maintenance.

**Disclosure.** The authors declare no conflicts of interest.

**Data availability.** Data underlying the results presented in this paper are not publicly available at this time but may be obtained from the authors upon reasonable request.